\documentclass[preprint,12pt]{elsarticle}

\usepackage[normalem]{ulem}  
\usepackage{color} 

\renewcommand\sout{\bgroup \color{red} \ULdepth=-.5ex \ULset}

\journal{Physics Letters B}

\usepackage{graphicx}
\usepackage{amssymb}
\usepackage{amsmath}
\usepackage{color}

\begin{document}

\begin{frontmatter}

\title{Charmed Tetraquarks $T_{cc}$ and $T_{cs}$\\ from Dynamical Lattice QCD Simulations}

\author[a]{Yoichi~Ikeda}
\ead{yikeda@riken.jp}
\author[b]{Bruno~Charron}
\author[c,d]{Sinya~Aoki}
\author[a]{Takumi~Doi}
\author[a,e]{Tetsuo~Hatsuda}
\author[f]{Takashi~Inoue}
\author[d]{Noriyoshi~Ishii}
\author[c]{Keiko~Murano}
\author[d]{Hidekatsu~Nemura}
\author[d]{Kenji~Sasaki}
\author{(HAL QCD Collaboration)}
\author{\includegraphics[width=.20\textwidth]{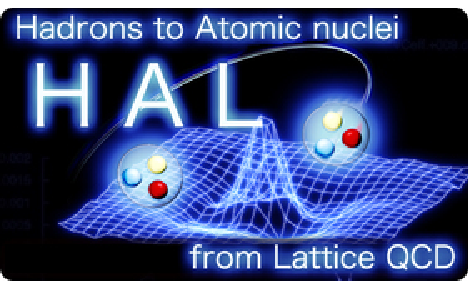}}
\address[a]{Theoretical Research Division,Nishina Center, RIKEN, Saitama 351-0198, Japan}
\address[b]{Department of Physics, The University of Tokyo, Tokyo 113-0033, Japan}
\address[c]{Yukawa Institute of Theoretical Physics, Kyoto University, Kyoto 606-8502, Japan}
\address[d]{Center for Computational Sciences, University of Tsukuba, Ibaraki 305-8571, Japan}
\address[e]{Kavli IPMU (WPI), The University of Tokyo, Chiba 606-8502, Japan}
\address[f]{College of Bioresource Science, Nihon University, Kanagawa 252-0880, Japan}

\begin{abstract}
Charmed tetraquarks $T_{cc}=(cc\bar{u}\bar{d})$ 
and $T_{cs}=(cs\bar{u}\bar{d})$ are studied through the S-wave meson-meson interactions,
$D$-$D$, $\bar{K}$-$D$, $D$-$D^{*}$ and $\bar{K}$-$D^{*}$, on the basis of the 
 (2+1)-flavor lattice QCD simulations 
 with the pion mass $m_{\pi} \simeq $410, 570 and 700 MeV.
 For the charm quark,  the relativistic heavy quark action is employed to treat its dynamics on the lattice.
 Using the HAL QCD method, we extract  the S-wave potentials in lattice QCD simulations, from which the meson-meson scattering phase shifts  are calculated.  The phase shifts in the  isospin triplet ($I$=1) channels 
 indicate repulsive interactions, while those in the $I=0$ channels suggest  attraction, growing as $m_{\pi}$ decreases. This is particularly prominent in the 
 $T_{cc} (J^P=1^+,I=0)$ channel, though
  neither bound state nor resonance are found in the range
  $m_{\pi} =410-700$ MeV. We make a qualitative comparison of our results with the
  phenomenological diquark picture.   
 \end{abstract}
\end{frontmatter}

\section{Introduction}
\label{intro}

One of the long standing challenges in hadron physics is to establish and classify
 genuine multiquark states other than baryons (3 quark states) and mesons (quark-antiquark states)~\cite{Jaffe1977}.
 In particular,
  charmed tetraquarks (such as $T_{cc}~(cc\bar{u}\bar{d})$, $T_{cs}~(cs\bar{u}\bar{d})$)
  and bottomed tetraquarks (such as $T_{bb}$, $T_{bc}$, $T_{bs}$)
 are unambiguous candidates for such multiquark states~\cite{Zouzou:1986qh,Lipkin1986,Carlson:1987hh,Manohar1993},
   since there are no annihilations among the four quarks.
  If they form bound states or resonances with respect to the corresponding
  two-meson thresholds,  
  they could be experimentally observed in  
  $B$-factories and relativistic heavy-ion colliders  
  \cite{DelFabbro:2004ta,Cho:2010db,Hyodo2012,Rashidin:2013voa}.

  In this Letter, we exclusively investigate the charmed tetraquarks $T_{cc}$ and $T_{cs}$.  
  To understand a possible reason why they may appear as bound states 
  below the two meson threshold,  let us consider the diquark picture
  \cite{Selem:2006nd} as a working hypothesis, where 
 $\bar{u}\bar{d}$ in the color {\bf 3}, spin-singlet ($S$=0), 
 isospin-singlet ($I$=0) channel 
  is  denoted as a ``good'' diquark,  due to  the large attraction between $\bar{u}$ and $\bar{d}$  generated through a gluon exchange. 
  We now assume  that good diquarks
   are the main substructure of charmed tetraquarks. 
  Then the color-singletness of hadrons and the quark Pauli principle
   constrain the   possible  low-mass tetraquarks as follows.
   \begin{enumerate}
  \item[(i)]
   $T_{cc}$ ($J^P=1^+$, $I=0$)  in which  $\bar{u}\bar{d}$
   forms a good diquark while the diquark $cc$ with  color ${\bf 3^*}$ and  $S=1$ has
   a weak repulsion between the two charm quarks.
   This state couples to the $D$-$D^{*}$ system.
  \item[(ii)] $T_{cs}$ ($J^P=1^+$, $I=0$)  in which  $\bar{u}\bar{d}$
   forms a good diquark while the diquark $cs$  with color ${\bf 3^*}$ and  $S=1$
   has a weak repulsion.
   This state couples to the $\bar{K}$-$D^{*}$ system.
  \item[(iii)]  $T_{cs}$ ($J^P=0^+$, $I=0$)  in which  $\bar{u}\bar{d}$
   forms a good diquark while the diquark $cs$ with color ${\bf 3^*}$ and  $S=0$ has an attraction.
   This state couples to the $\bar{K}$-$D$ system.
  \end{enumerate}
   
 Quantitatively, however,  predictions for the binding energies of charmed tetraquarks widely spread, ranging from 
   negative values (resonance) to 100 MeV (deeply bound)  with respect to the two-meson thresholds, 
  depending on the details of the dynamical models (diquark model,
   dynamical four-body calculation in the constituent quark model,
  meson-meson molecular  model, lattice QCD in the heavy-quark limit, etc)
  \cite{Yasui2009,Carames:2011zz,Ohkoda2012,Vijande:2013qr,Brown:2012tm,Bicudo:2012qt}.
 Therefore, a quantitative prediction for charmed tetraquarks requires
 a careful study in full lattice QCD with a finite charm-quark mass
  \footnote{The importance of the finite charm-quark mass to extract the
   $c$-$\bar{c}$ potentials from lattice QCD was reported previously in
     \cite{Ikeda:2011bs,Kawanai:2011xb}}.
     
  In this Letter, we report our first results  on  the interactions in the 
  $D$-$D$, $\bar{K}$-$D$, $D$-$D^{*}$ and $\bar{K}$-$D^{*}$ systems in
  (2+1)-flavor lattice QCD simulations.
   The dynamics of the charm quarks are incorporated on the lattice 
    with the relativistic heavy quark action~\cite{RHQ_Aoki}.
   The meson-meson scattering phase shifts are derived from the corresponding potentials calculated 
  on the lattice by the HAL QCD method ~\cite{Ishii2007a,Aoki2010,Ishii2012} 
   (reviewed in \cite{HAL2012}), which was recently shown to be  
  quite  accurate to describe some meson-meson scattering phase shifts \cite{Kurth:2013tua}.
   
This Letter is organized as follows.
In Sec.~2, we present the HAL QCD method to extract the potential between two mesons. We then show the numerical setup of our lattice QCD simulations in Sec.~3.
In Sec.~4, we show our numerical results for the potentials, from which scattering phase shifts and  scattering lengths  are extracted for three different quark masses. 
Sec.~5 is devoted to a summary and a discussion.

\section{HAL QCD method for the meson-meson interaction}

In QCD, the two-meson correlation function can be expanded as
\begin{eqnarray}
F (\vec r, t) &\equiv& 
\sum_{\vec x} \left\langle 0 \right| 
\mathcal{O}_{h_1}(\vec x + \vec r, t) \mathcal{O}_{h_2}(\vec x, t) \overline{{\cal J}}_{h_1 h_2}(t=0)
\left| 0 \right\rangle \nonumber \\
 &=& 
\sum_{\vec x, n}A_n \left\langle 0 \right|
\mathcal{O}_{h_1}(\vec x+\vec r, t) \mathcal{O}_{h_2}(\vec x, t)
\left| n \right\rangle e^{-W_n t} +\dots,
\label{4-point}
\end{eqnarray}
with 
$A_n = 
\left\langle n \right| \overline{{\cal J}}_{h_1 h_2}(t=0) \left| 0 \right\rangle,
$
$\overline{{\cal J}}_{h_1 h_2}(t=0)$ stands for a source operator at $t_{\rm src.}=0$ which creates
two meson states and  $\mathcal{O}_{h_{1,2}}$ is a point-like interpolating sink operator for the hadron $h_{1,2}$.
$W_n = \sqrt{m_1^2 + \vec k_n^2}+\sqrt{m_2^2 + \vec k_n^2}$ is the relativistic energy
of the $n$-th eigenstate $\left| n \right\rangle$ for two mesons, and ellipses represent inelastic contributions.

Consider  $t$ sufficiently larger than $ t_{\rm src.} $ that  
the contributions from elastic scattering states and possible bound states remain while 
those from inelastic states become negligible. Then, 
Eq.~(\ref{4-point}) becomes
$F (\vec r, t)  \rightarrow \sum_n A_n \phi_{n}(\vec r) e^{-W_n t}$,
where $\phi_{n}(\vec r)$ is an equal-time Nambu-Bethe-Salpeter (NBS) 
wave function~\cite{Miransky}, 
from which the HAL QCD potential $U$ is defined~\cite{Aoki2010} as a solution of
\begin{equation}
H_0 \phi_n(\vec{r}) 
+ \int d\vec{r}' U(\vec{r},\vec{r'}) \phi_n(\vec{r'})
= E_n \phi_n(\vec{r}),
\label{eff_Sch_eq}
\end{equation}
for all elastic eigenstates $n$, where  $H_0 = -\nabla^2/2\mu $ 
 with $\mu=m_1 m_2 /(m_1 + m_2)$ and $E_n=\vec k_n^2 / 2\mu$ is
  a kinetic energy.
  Here the non-local but $n$-independent potential  $U(\vec{r}, \vec{r'})$ can be shown to exist, by explicitly constructing it as
\begin{equation}
U(\vec{r},\vec{r'}) 
= \sum_n (E_n - H_0) \phi_n(\vec{r})\cdot  \tilde{\phi}_n^* (\vec r'), 
\label{non-localU}
\end{equation}
where
 $\tilde{\phi}_n^* (\vec r')$ is the dual basis associated with ${\phi}_n (\vec r')$, and the summation over $n$ is restricted to elastic channels. (For details and proofs, see  \cite{Aoki2010,HAL2012}).
 
In principle, the potential is extracted from $F(\vec{r},t)$ at large $t$, when it is dominated by the $n=0$ state (i.e. the ground state) contribution~\cite{Aoki2010,HAL2012}. 
In practice, however, $F(\vec{r},t)$ is usually noisier at larger $t$, so that an accurate determination of potentials in this way becomes difficult. 

To overcome this practical difficulty,  an alternative method has been proposed in~\cite{Ishii2012}. 
Since $U(\vec{r}, \vec{r'})$ is $n$-independent by definition,  a normalized correlation function $R(\vec{r},t) = F(\vec{r},t)/e^{-(m_1+m_2) t}$ satisfies
\begin{eqnarray}
\label{t-dep_schrod}
&&\biggl( -\frac{\partial}{\partial t} - H_0 \biggr) R(\vec r, t) = \sum_n A_n ( \Delta W_n -H_0)\phi_n(\vec r)e^{-\Delta W_nt}   \nonumber \\
&\simeq& \sum_n  A_n (E_n - H_0)\phi_n(\vec{r}) e^{-\Delta W_nt} 
= \int d \vec r' ~ U(\vec r, \vec r') R(\vec r', t) ,
\end{eqnarray}
where the non-relativistic approximation that 
$\Delta W_n \equiv W_n - m_1-m_2 = E_n + O(k^4/m_1^3, k^4/m_2^3)$ is used
\footnote{This approximation can be avoided if we allow higher order time derivatives
in Eq.~(\ref{t-dep_schrod}),
whose contribution, however, turns out to be numerically negligible
for the systems investigated in this Letter.}.
In the velocity expansion of the non-local potential, we finally obtain
the leading order potential as
\begin{eqnarray}
V_{\rm LO}(\vec r) =
-\frac{(\partial/\partial t)R(\vec r, t)}{R(\vec r, t)} 
-\frac{H_0 R(\vec r, t)}{R(\vec r, t)},
\label{t-dep_local}
\end{eqnarray}
within the non-relativistic approximation.

To extract S-wave potentials on the lattice, we consider the projection of the normalized correlation function on the $A_1^+$ representation of the cubic group (containing the $J=0$ representation of the rotational group)
\begin{equation}
R(\vec r,t; A_1^+) \equiv P^{(A_1^+)} R( \vec{r}, t)
=
\frac{1}{24}\sum_{g \in O}  \chi^{(A_1^+)}(g)R(g^{-1} \vec{r}, t),
\label{s-wave_A1}
\end{equation}
where $g \in O$ are elements of the cubic group, and $\chi^{(A_1)}(g)(\equiv 1)$
are the associated characters of the $A_1$ representation.

\section{Numerical setup}

We employ (2+1)-flavor full QCD gauge configurations
 generated by the PACS-CS collaboration~\cite{PACS-CS2009,PACS-CS2010} on a $32^3 \times 64$ lattice
with the renormalization group improved gauge action at $\beta = 1.90$ and 
the non-perturbatively $O(a)$-improved Wilson quark action ($C_{\rm SW}=1.715$) 
at $(\kappa_{ud}, \kappa_s) =$ $(0.13754,0.13640)$, $(0.13727, 0.13640)$, and $(0.13700,0.13640)$.
These parameters correspond to the lattice 
 cutoff $a^{-1} = 2176$ MeV (lattice spacing $a = 0.0907(13)$ fm),  determined from $\pi$, $K$ and $\Omega$ masses as inputs~\cite{PACS-CS2009}, leading to the spatial lattice 
 volume  $L^3 \simeq (2.9 {\rm fm})^3$.

As for the charm quark, we employ a relativistic heavy quark (RHQ) action proposed in Ref.~\cite{RHQ_Aoki},
which is designed to remove the leading and next-to-leading order cutoff errors
 associated with heavy quark mass, ${\cal O}((m_Q a)^{n})$ 
 and ${\cal O}((m_Q a)^{n} (a \Lambda_{QCD}))$, respectively.
The RHQ action is given by
\begin{eqnarray}
S_Q &=& \sum_{x,y} \bar{Q}(x) D(x,y) Q(y)  , \\
D(x,y) &=& \delta_{x,y} 
- \kappa_{Q} \sum_{i=1}^{3} \bigl[ 
( r_s - \nu \gamma_i ) U_{x,i} \delta_{x+\hat{i},y} +
( r_s + \nu \gamma_i ) U^{\dagger}_{x,i} \delta_{x,y+\hat{i}} \bigr] \nonumber \\
& & 
- \kappa_{Q} \bigl[ 
( r_t - \nu \gamma_4 ) U_{x,4} \delta_{x+\hat{4},y} +
( r_t + \nu \gamma_4 ) U^{\dagger}_{x,4} \delta_{x,y+\hat{4}} \bigr] \nonumber \\
& & 
- \kappa_{Q} \biggl[ 
c_{B} \sum_{i,j} F_{ij} \sigma_{ij} + c_{E} \sum_{i} F_{i4} \sigma_{i4} 
\biggl] \delta_{x,y}  .
\end{eqnarray}
Parameters of the action are $\kappa_Q$, $r_s$, $r_t$, $\nu$, $c_{B}$ and $c_{E}$,
while the redundant parameter $r_t$ is chosen to be $ 1$.
In our simulations, we take the same parameters as in Ref.~\cite{Namekawa2011},
where the 1S charmonium mass and its relativistic dispersion relation are reproduced.
The RHQ parameters are summarized in Table~\ref{tab:RHQ_param}.

\begin{table}[htbp]
   \centering
   \begin{tabular}{ccccc}
      \hline
      \hline
$\kappa_{Q}$ & $r_s$ & $\nu$ & $c_{B}$ & $c_{E}$ \\
      \hline 
0.10959947  & 1.1881607 & 1.1450511 & 1.9849139 & 1.7819512 \\
      \hline 
      \hline
   \end{tabular}
   \caption{ Parameters of the RHQ action in our calculations. See Ref.~\cite{Namekawa2011} for more details.
   }
   \label{tab:RHQ_param}
\end{table}

Periodic boundary conditions are imposed in the three spacial directions, 
while Dirichlet boundary conditions are taken for the temporal direction at $t/a=\pm 32$
to avoid contaminations from the opposite propagation of mesons in time.
Throughout this study, we employ  local interpolating operators for mesons,
$\phi(x) = \bar{q}(x) \Gamma q(x)$, where $\Gamma$ denotes a $4\times 4$ matrix acting on spinor indices.
We take $\Gamma = \gamma_5$ for  pseudo-scalar mesons ($D$ and $\bar{K}$)
and $\Gamma = \gamma_i$ for vector mesons ($D^*$).
Meson masses calculated in this work are listed in Table~\ref{tab1}
together with the number of configurations used in this work.
We measure the correlation function in Eq.~(\ref{4-point}) with a source at one time-slice for each configuration,
and the forward and backward propagations are averaged to enhance the statistics.
We have checked that the dispersion relation of the 1S charmonium state at our heaviest pion mass, $m_{\pi} \sim 700$MeV,  gives a reasonable value of the effective speed of light, 
$c_{\rm eff} = 0.987(2)$.

\begin{table}[htbp]
   \centering
   \begin{tabular}{cccc}
      \hline
      \hline
$(\kappa_{ud}, \kappa_s)$ & $(0.13754,0.13640)$ & $(0.13727, 0.13640)$ & $(0.13700,0.13640)$\\
      \hline 
confs. & 450 & 400 & 399 \\
      \hline 
$m_{\pi}$ (MeV) & 411(2) & 572(2) & 699(1) \\
      \hline 
$m_{K}$ (MeV)   & 635(2) & 714(1) & 787(1) \\
      \hline 
$m_{\eta_c}$ (MeV)   & 2988(2) & 3005(1) & 3024(1) \\
      \hline
$m_{J/ \Psi}$ (MeV)   & 3097(2) & 3118(1) & 3142(1) \\
      \hline
$m_{D}$ (MeV)   & 1902(3) & 1946(1) & 1999(1) \\
      \hline
$m_{D^*}$ (MeV)   & 2048(12) & 2099(6) & 2159(4) \\
      \hline 
      \hline
   \end{tabular}
   \caption{ Meson masses obtained in this study.
   We employ the scale determined in Ref.~\cite{PACS-CS2009}. }
   \label{tab1}
\end{table}

As for the source operators of the $D$-$D$, $D$-$D^*$, $\bar{K}$-$D$ and $\bar{K}$-$D^*$
in isospin $I$ channels,
we take the following wall sources:
\begin{eqnarray}
\overline{{\cal J}}_{D D}(t=0) &=& 
\sum_{\vec x_1, \vec x_2, \vec x_3, \vec x_4}
[\bar{c}(\vec x_1, t) \gamma_5 u(\vec x_2, t)
 \bar{c}(\vec x_3, t) \gamma_5 d(\vec x_4, t) \nonumber \\
& & + (-)^{I+1}
 \bar{c}(\vec x_1, t) \gamma_5 d(\vec x_2, t)
 \bar{c}(\vec x_3, t) \gamma_5 u(\vec x_4, t) ] , \\
\overline{{\cal J}}_{D D^*}(t=0) &=& 
\sum_{\vec x_1, \vec x_2, \vec x_3, \vec x_4}
[\bar{c}(\vec x_1, t) \gamma_5 u(\vec x_2, t)
 \bar{c}(\vec x_3, t) \gamma_i d(\vec x_4, t) \nonumber \\
& & + (-)^{I+1}
 \bar{c}(\vec x_1, t) \gamma_5 d(\vec x_2, t)
 \bar{c}(\vec x_3, t) \gamma_i u(\vec x_4, t) ] , \\
\overline{{\cal J}}_{\bar{K} D}(t=0) &=& 
\sum_{\vec x_1, \vec x_2, \vec x_3, \vec x_4}
[\bar{s}(\vec x_1, t) \gamma_5 u(\vec x_2, t)
 \bar{c}(\vec x_3, t) \gamma_5 d(\vec x_4, t) \nonumber \\
& & + (-)^{I+1}
 \bar{s}(\vec x_1, t) \gamma_5 d(\vec x_2, t)
 \bar{c}(\vec x_3, t) \gamma_5 u(\vec x_4, t) ] , \\
\overline{{\cal J}}_{\bar{K} D^*}(t=0) &=& 
\sum_{\vec x_1, \vec x_2, \vec x_3, \vec x_4}
[\bar{s}(\vec x_1, t) \gamma_5 u(\vec x_2, t)
 \bar{c}(\vec x_3, t) \gamma_i d(\vec x_4, t) \nonumber \\
& & + (-)^{I+1}
 \bar{s}(\vec x_1, t) \gamma_5 d(\vec x_2, t)
 \bar{c}(\vec x_3, t) \gamma_i u(\vec x_4, t) ] .
\end{eqnarray}

\section{Meson-meson potentials and scattering phase shifts}

With the above setup, we study the S-wave meson-meson interactions in the following channels related to $T_{cc}$ and $T_{cs}$
 with $J^P=0^+, 1^+$:
 $D$-$D$ ($J^{P}=0^{+},~I=1$), $\bar{K}$-$D$ ($J^{P}=0^{+},~I=0,1$),
 $D$-$D^{*}$ ($J^{P}=1^{+},~I=0,1$),
 and $\bar{K}$-$D^{*}$ ($J^{P}=1^{+},~I=0,1$).
We show the potentials calculated from Eq.~(\ref{t-dep_local}) at the time-slice $t/a=16$.
Since the energy differences between the elastic and inelastic thresholds
are 200-300MeV as shown in Table~\ref{tab_th}, 
the inelastic contributions in Eq.~(\ref{4-point}) are expected to be suppressed in the time-slices $t/a > 11$.
\begin{table}[htbp]
   \centering
   \begin{tabular}{cccc}
      \hline
      \hline
Threshold energies & $m_{\pi}$=411(2) & $m_{\pi}$=572(2) & $m_{\pi}$=699(1) \\
      \hline 
$E_{DD}$  & 3805(5) & 3893(3) & 3999(3) \\
$E_{D^* D^*}$  & 4097(23) & 4199(11) & 4319(7) \\
      \hline 
$E_{DD^*}$  & 3951(12) & 4046(6) & 4159(5) \\
$E_{D^* D^*}$  & 4097(23) & 4199(11) & 4319(7) \\
      \hline
$E_{\bar{K}D}$  & 2538(3) & 2660(2) & 2785(2) \\
$E_{\bar{K}^*D^*}$  & 3075(15) & 3184(9) & 3314(8) \\
      \hline
$E_{\bar{K}D^*}$  & 2684(12) & 2814(6) & 2946(4) \\
$E_{\bar{K}^*D}$  & 2930(9) & 3031(7) & 3153(6) \\
      \hline 
      \hline
   \end{tabular}
   \caption{ The lowest (inelastic) threshold energies 
   for the $D$-$D$ ($D^*$-$D^*$), $D$-$D^*$ ($D^*$-$D^*$), $\bar{K}$-$D$ ($\bar{K}^*$-$D^*$) and 
   $\bar{K}$-$D^*$ ($\bar{K}^*$-$D$) in MeV unit.
   }
   \label{tab_th}
\end{table}

\begin{figure}[htb]
\includegraphics[width=0.45\textwidth,clip]{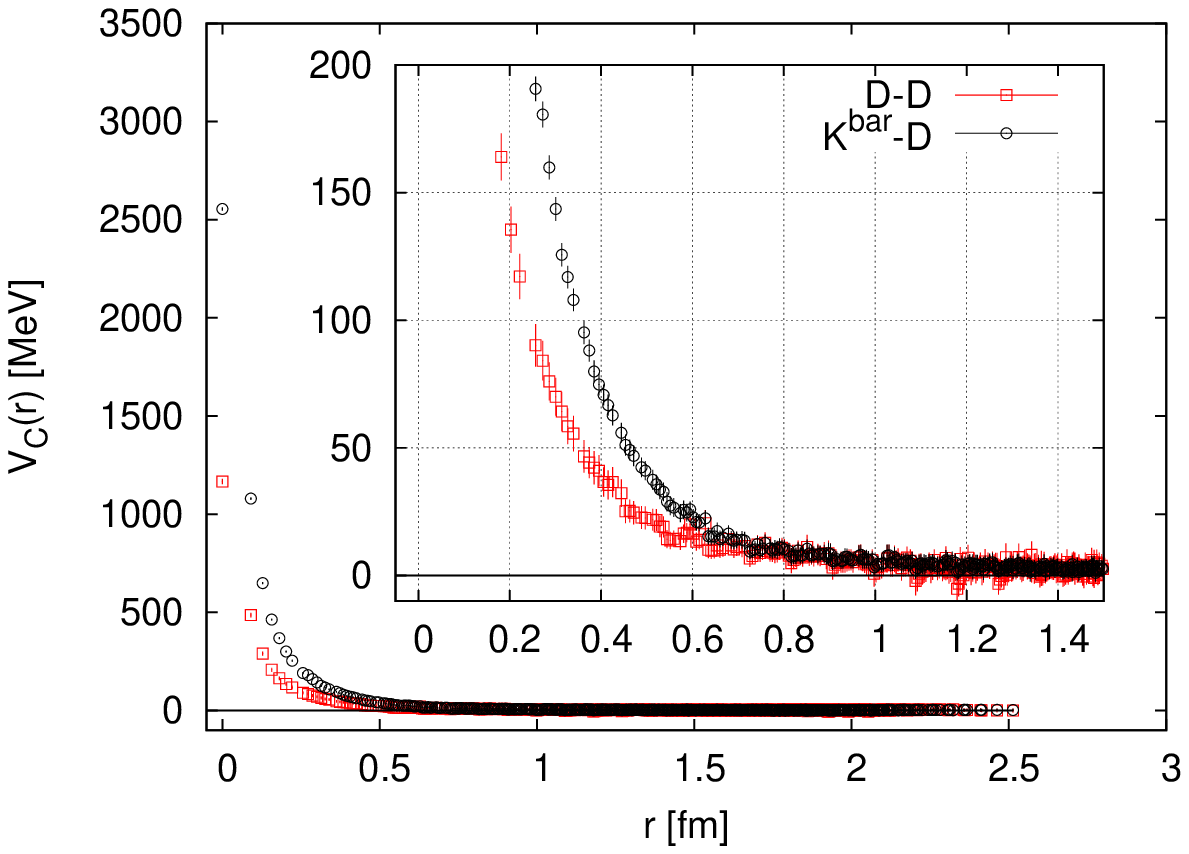}
\put(-140,120){{\scriptsize (a)}}\vspace{0.75cm}
\includegraphics[width=0.45\textwidth,clip]{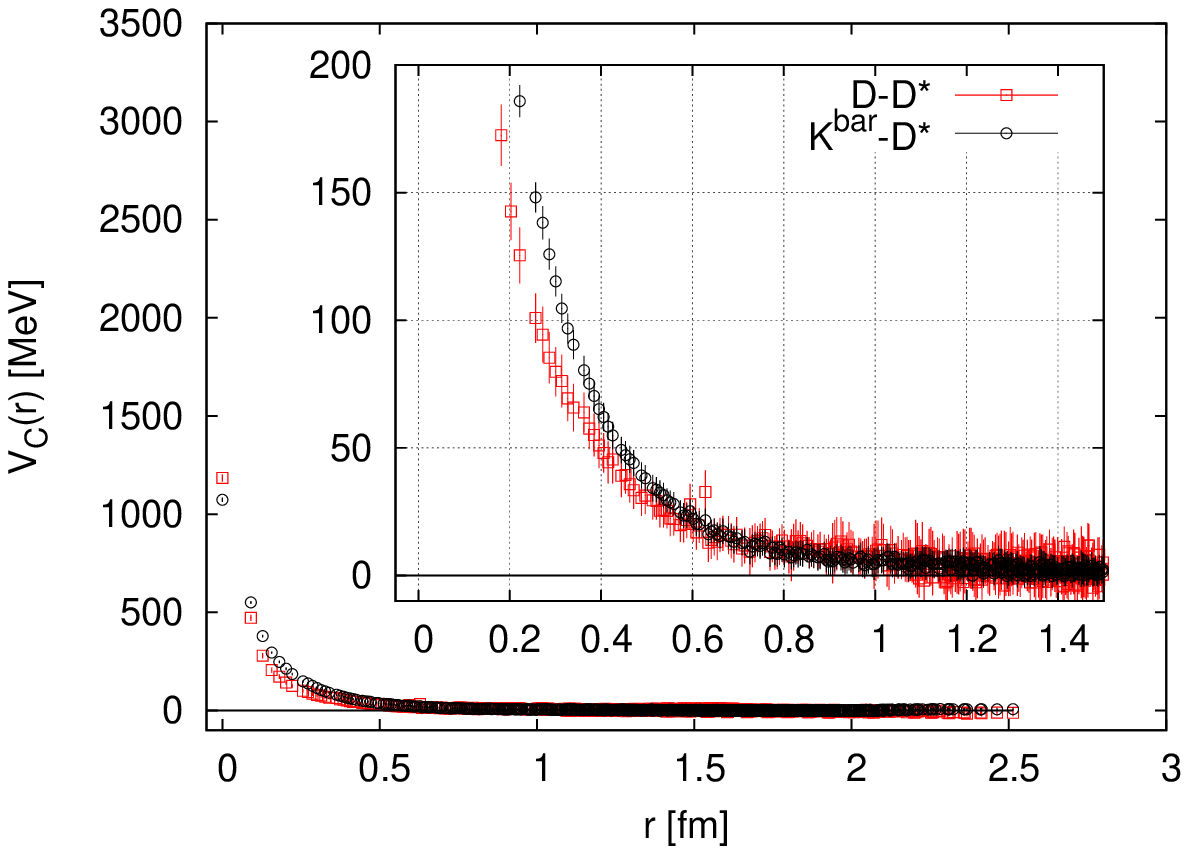}\\
\includegraphics[width=0.45\textwidth,clip]{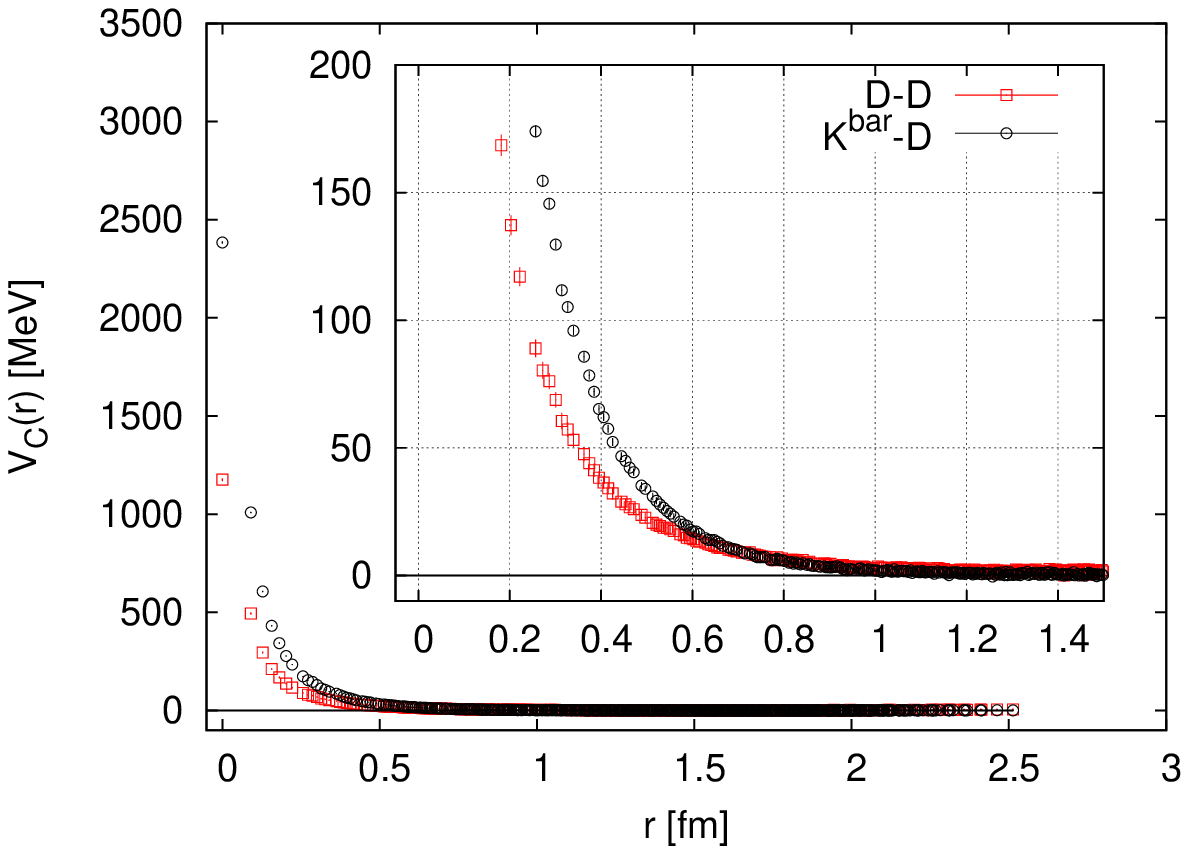}
\put(-140,120){{\scriptsize (b)}}\vspace{0.75cm}
\includegraphics[width=0.45\textwidth,clip]{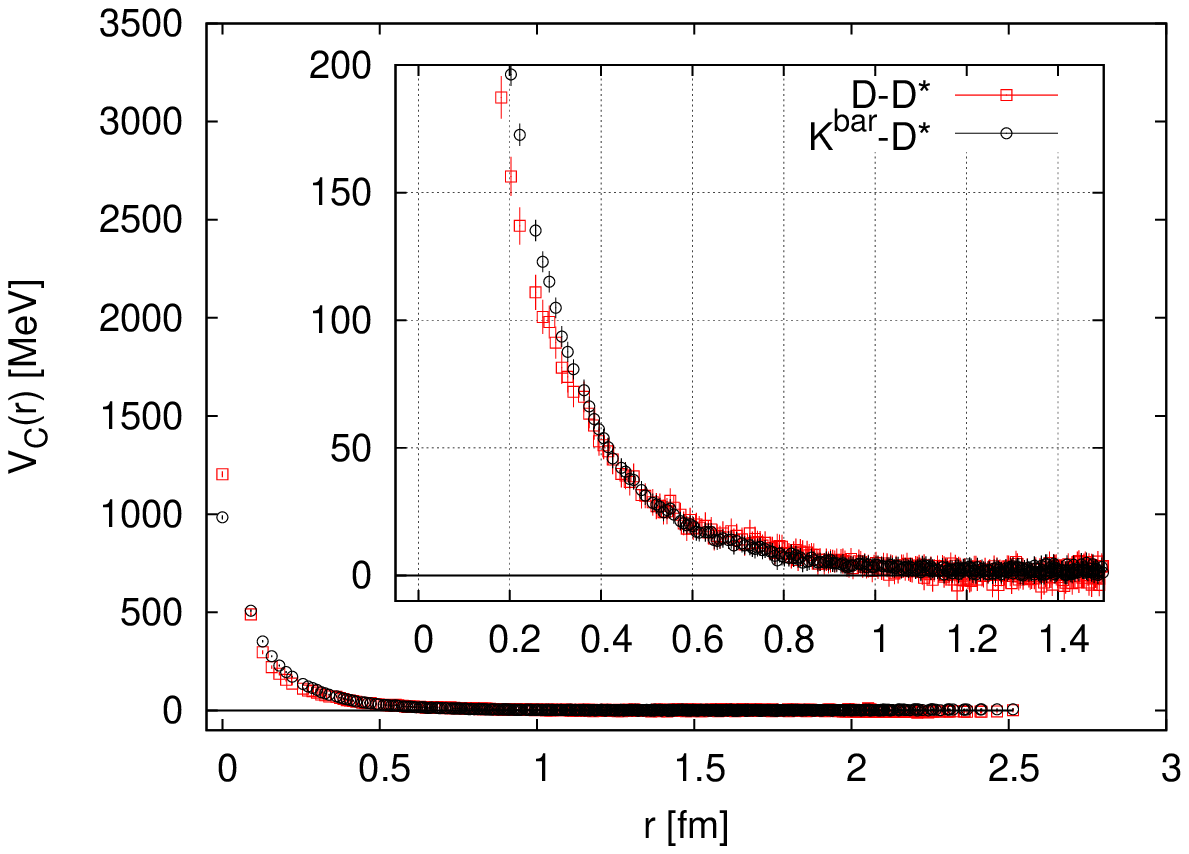}\\
\includegraphics[width=0.45\textwidth,clip]{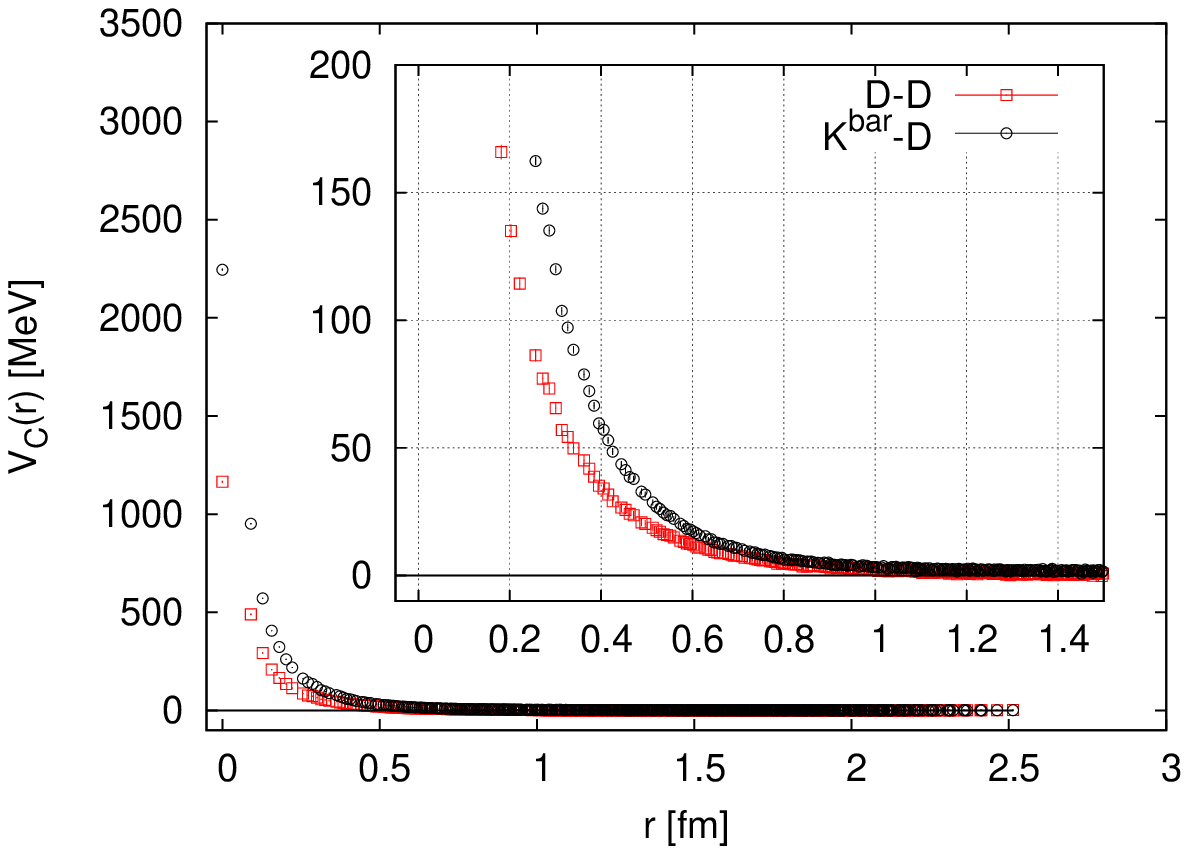}
\put(-140,120){{\scriptsize (c)}}\vspace{0.75cm}
\includegraphics[width=0.45\textwidth,clip]{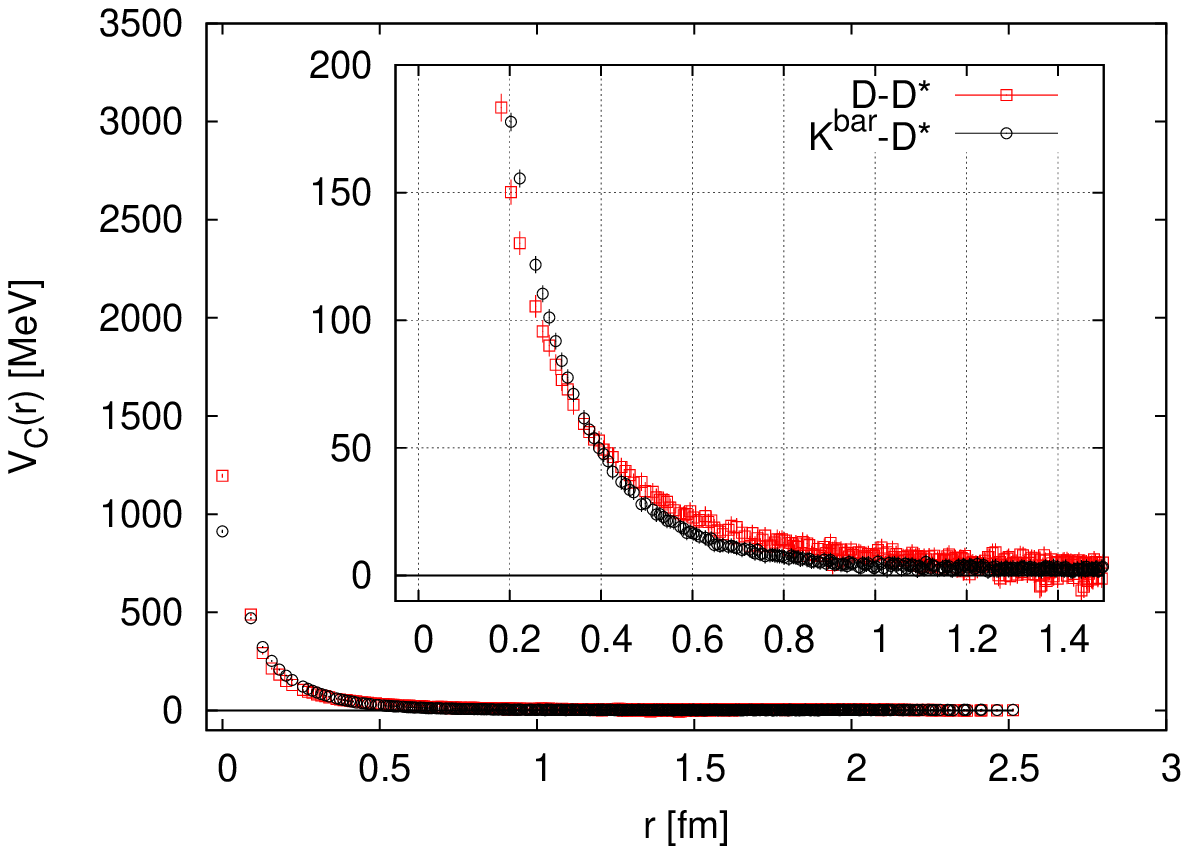}
\caption{(color online) Left three panels for the 
S-wave central potentials in the $D$-$D$ (square) and $\bar{K}$-$D$ (circle) 
channels with $(J^{P},I)=(0^{+},1)$.
Right three panels for 
the S-wave central potentials in the $D$-$D^{*}$ (square) and $\bar{K}$-$D^{*}$ (circle) 
channels with $(J^{P},I)=(1^{+},1)$.
(a), (b) and (c) are obtained at  $m_{\pi}\simeq 410$ MeV, $m_{\pi}\simeq 570$ MeV 
and  $m_{\pi}\simeq 700$ MeV, respectively.
 }
\label{fig1}
\end{figure}

In Fig.~\ref{fig1},  we show our results for the S-wave meson-meson potentials
in the $I=1$ channels with $J^{P}=0^{+}$ (left panels) and $J^{P}=1^{+}$ (right panels).
We find that all the potentials in the  $I=1$ channel are repulsive at all distances. This observation is consistent with the absence of good
$\bar{u}\bar{d}$ diquarks in the $I=1$ channel, as discussed in Section~\ref{intro}. 
Since the quark mass dependence of the potentials is rather weak,
it is unlikely that interactions in these channels turn into strong attractions to form bound states even at the physical quark mass. 
Phenomenological models also predict the absence of bound states in the $I=1$ channels. 

\begin{figure}[htb]
\includegraphics[width=0.45\textwidth,clip]{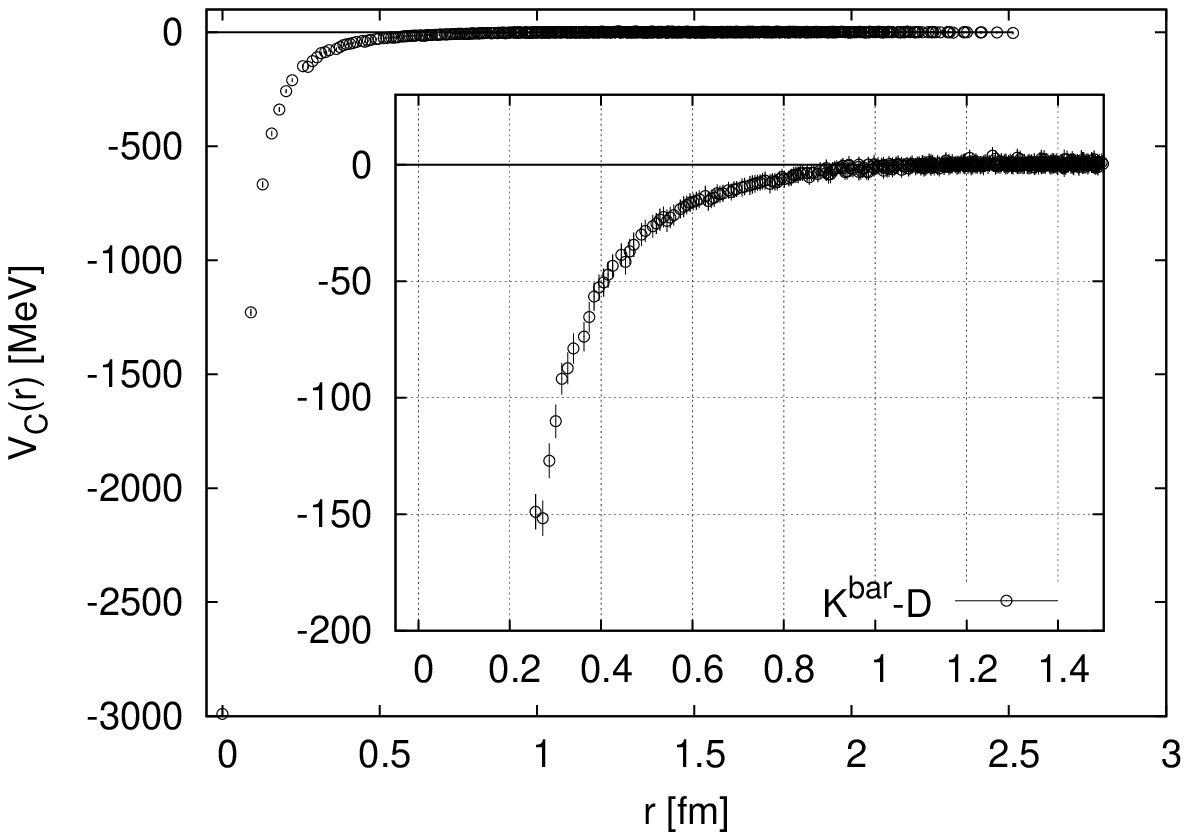}
\put(-140,120){{\scriptsize (a)}}\vspace{0.75cm}
\includegraphics[width=0.45\textwidth,clip]{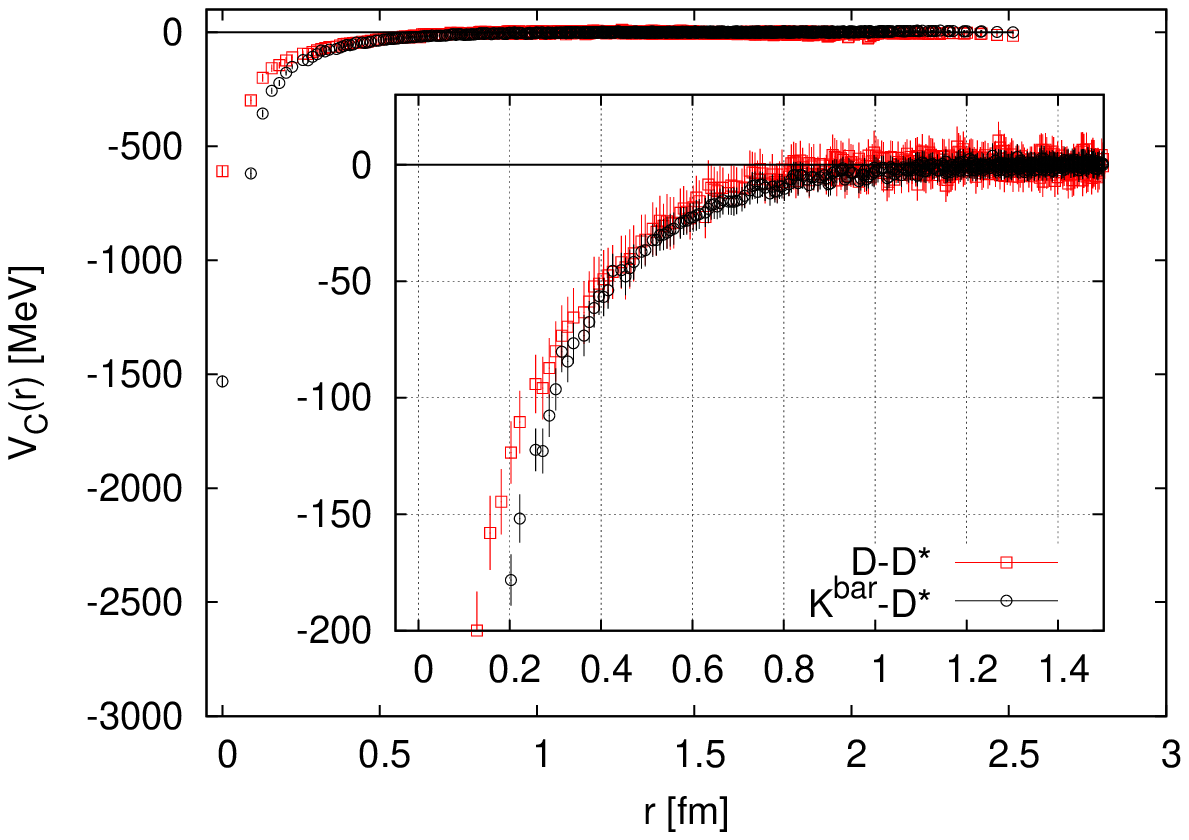}\\
\includegraphics[width=0.45\textwidth,clip]{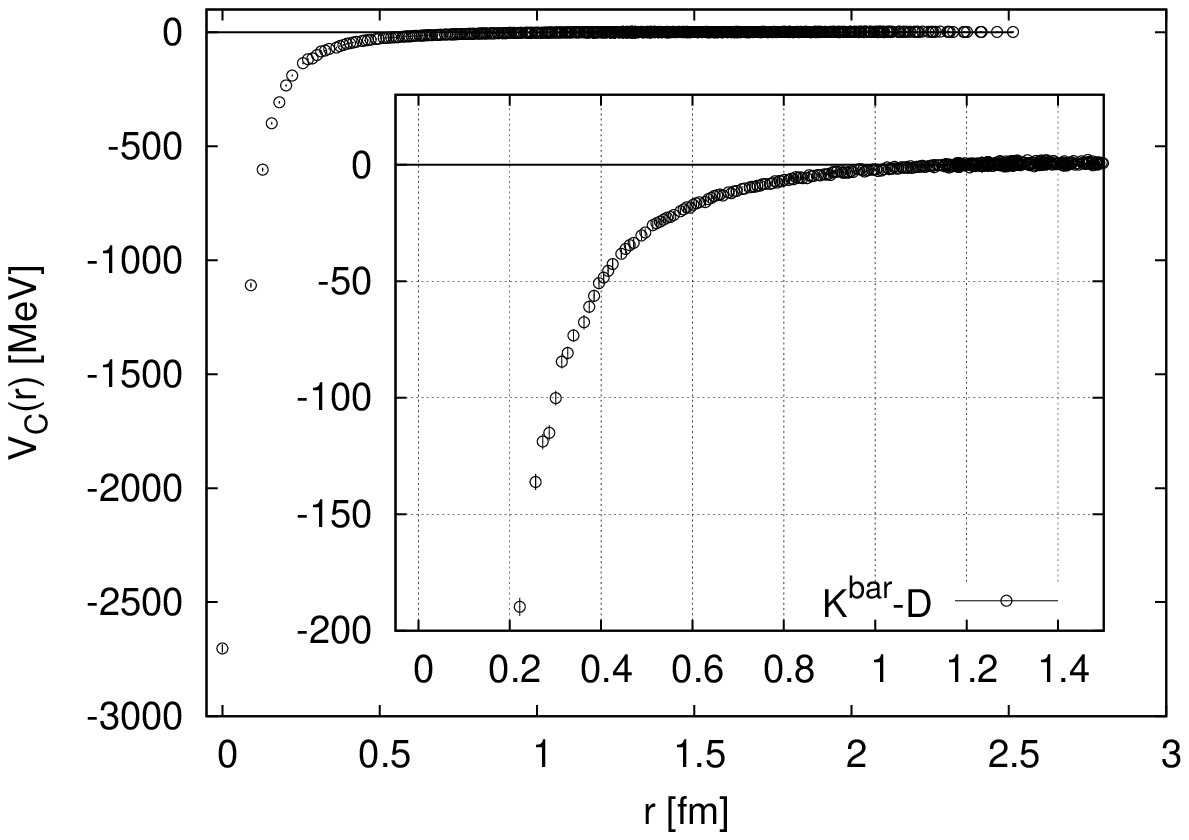}
\put(-140,120){{\scriptsize (b)}}\vspace{0.75cm}
\includegraphics[width=0.45\textwidth,clip]{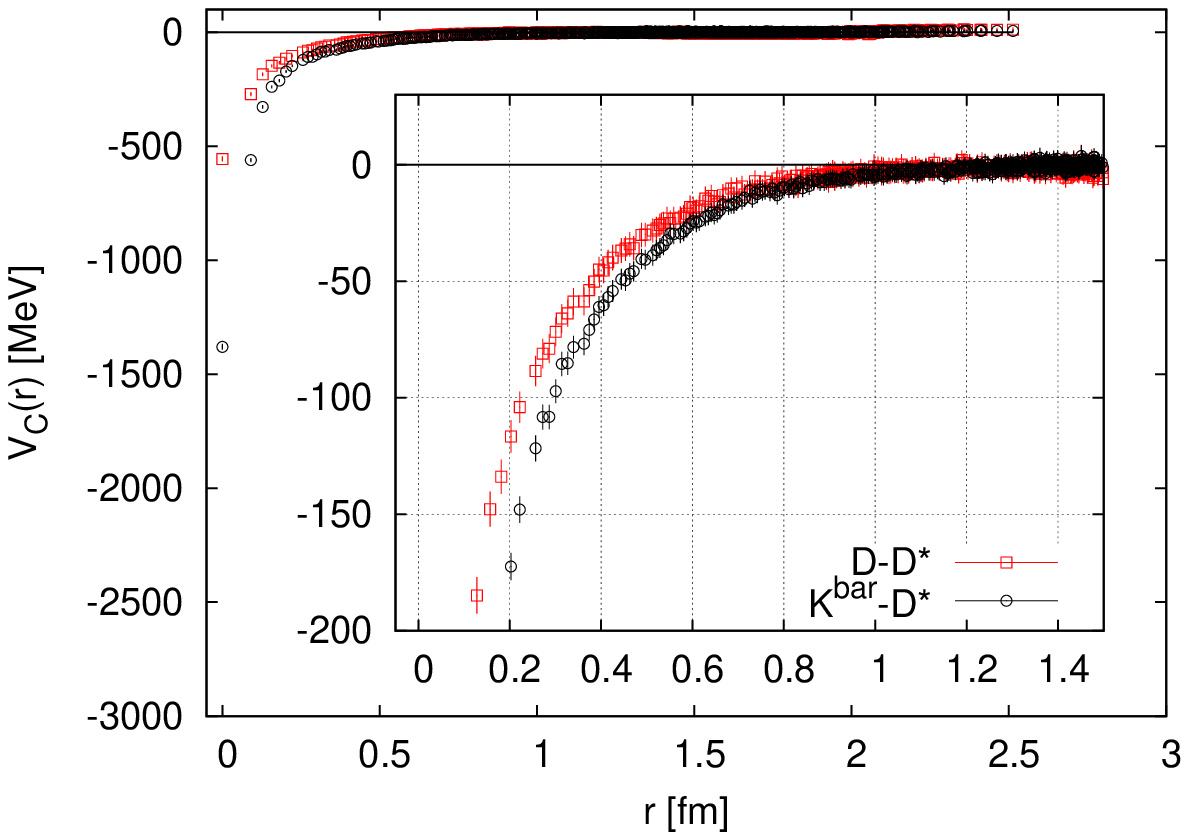}\\
\includegraphics[width=0.45\textwidth,clip]{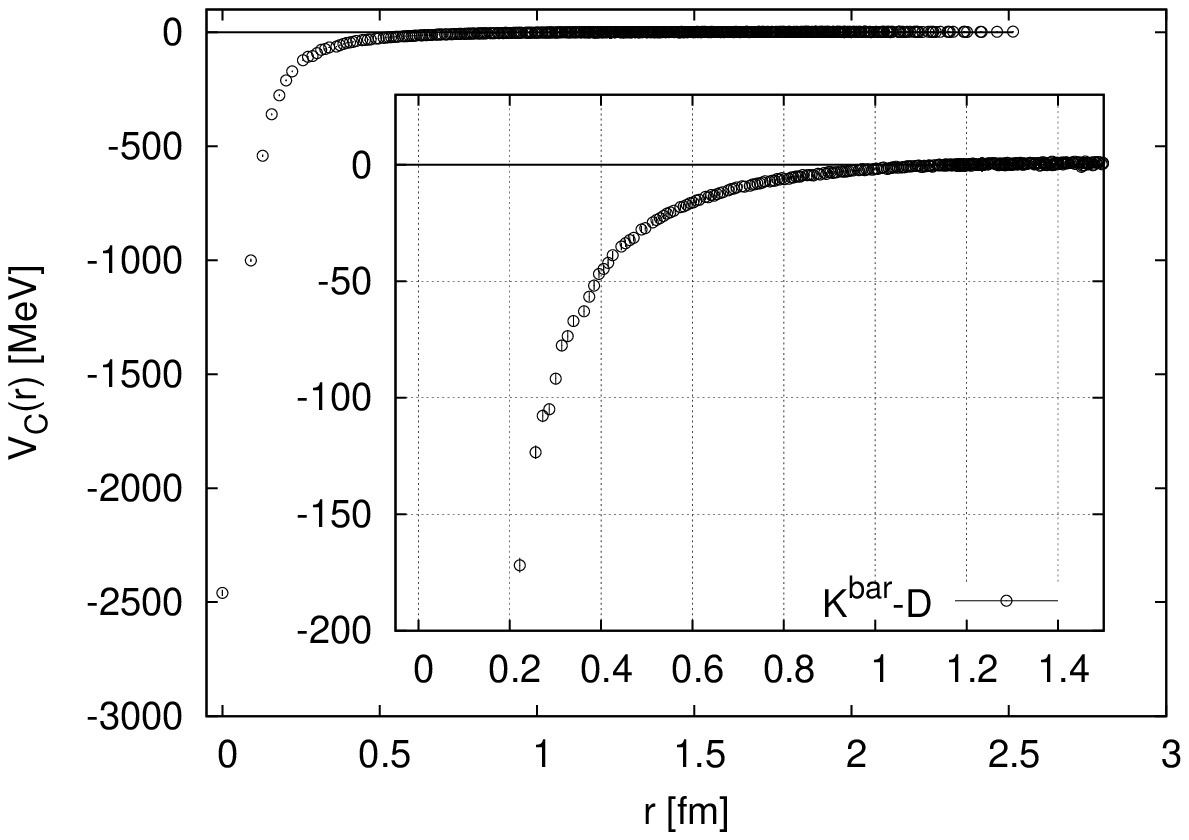}
\put(-140,120){{\scriptsize (c)}}\vspace{0.75cm}
\includegraphics[width=0.45\textwidth,clip]{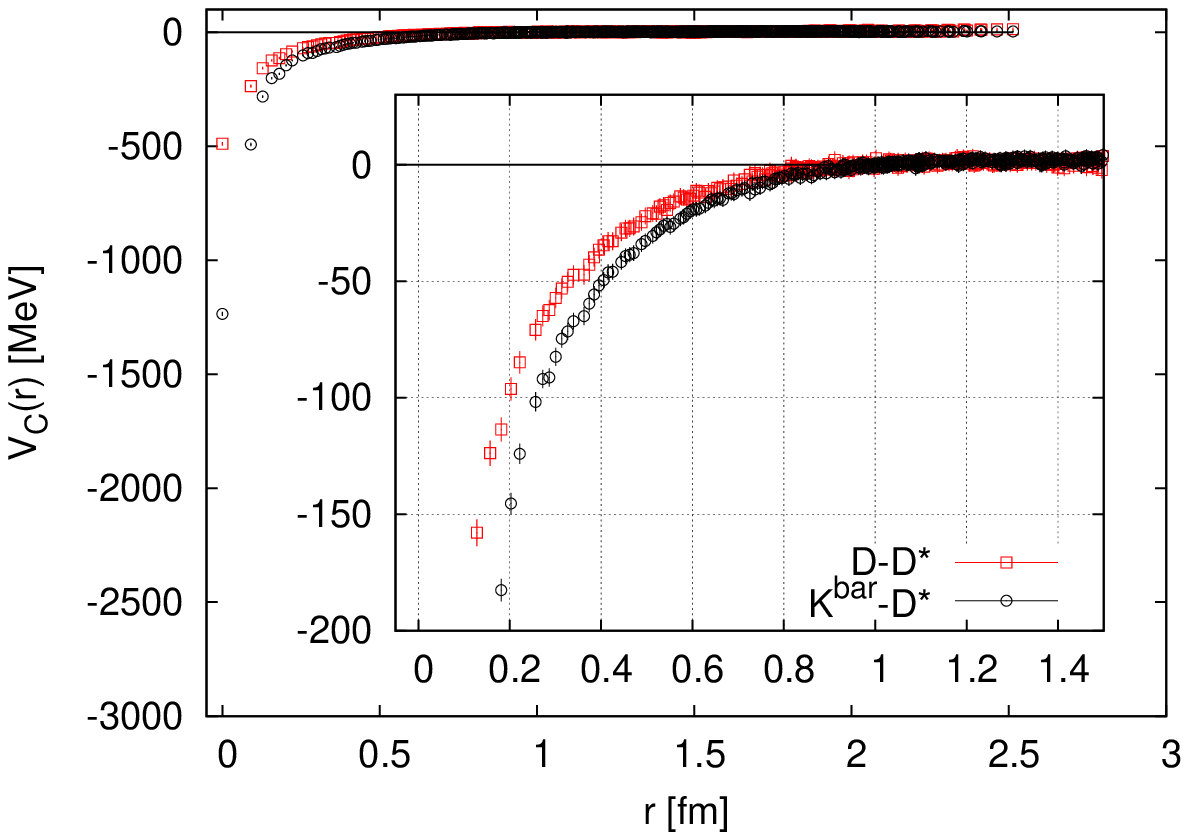}
\caption{(color online) Left three panels for 
the S-wave 
 central potentials in the $\bar{K}$-$D$ channel with $(J^{P},I)=(0^{+},0)$.
 Right three panels  
 for the S-wave central potentials in the $D$-$D^{*}$ (square) and $\bar{K}$-$D^{*}$ (circle) channels with
   $(J^{P},I)=(1^{+},0)$.
 (a), (b) and (c) are obtained at  $m_{\pi}\simeq 410$ MeV, $m_{\pi}\simeq 570$ MeV 
 and  $m_{\pi}\simeq 700$ MeV, respectively.
  }
\label{fig2}
\end{figure}

 In Fig.~\ref{fig2}, we show our results for the S-wave meson-meson central potentials
in the $I=0$ channels with $J^{P}=0^{+}$ (left panels)
 and $J^{P}=1^{+}$ (right panels). 
 Contrary to the previous results in the $I=1$ channels, 
 all the potentials in the $I=0$ channels show attractions at all distances without repulsive core. 
In addition, we find that all potentials become more attractive as the pion mass decreases,
and the attraction at short distance ($r = 0.2 \sim 0.3$fm) in the $\bar{K}$-$D$ channel is stronger than
in the $D$-$D^{*}$ and $\bar{K}$-$D^{*}$ channels.
Such tendency is again consistent with the existence of good $\bar{u}\bar{d}$ diquarks 
in the $I=0$ channel, as discussed in Section~\ref{intro}. 

\begin{figure}[htbp]
\begin{center}
\includegraphics[width=0.6\textwidth]{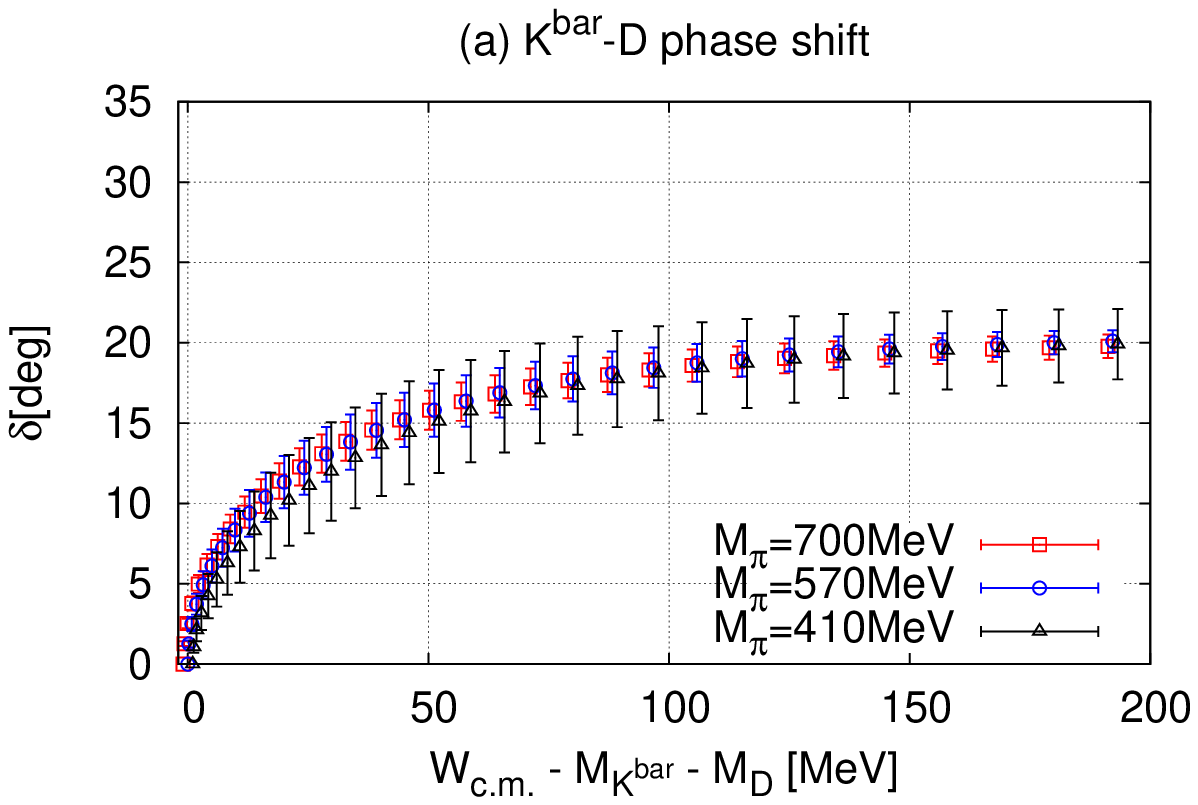}
\includegraphics[width=0.6\textwidth]{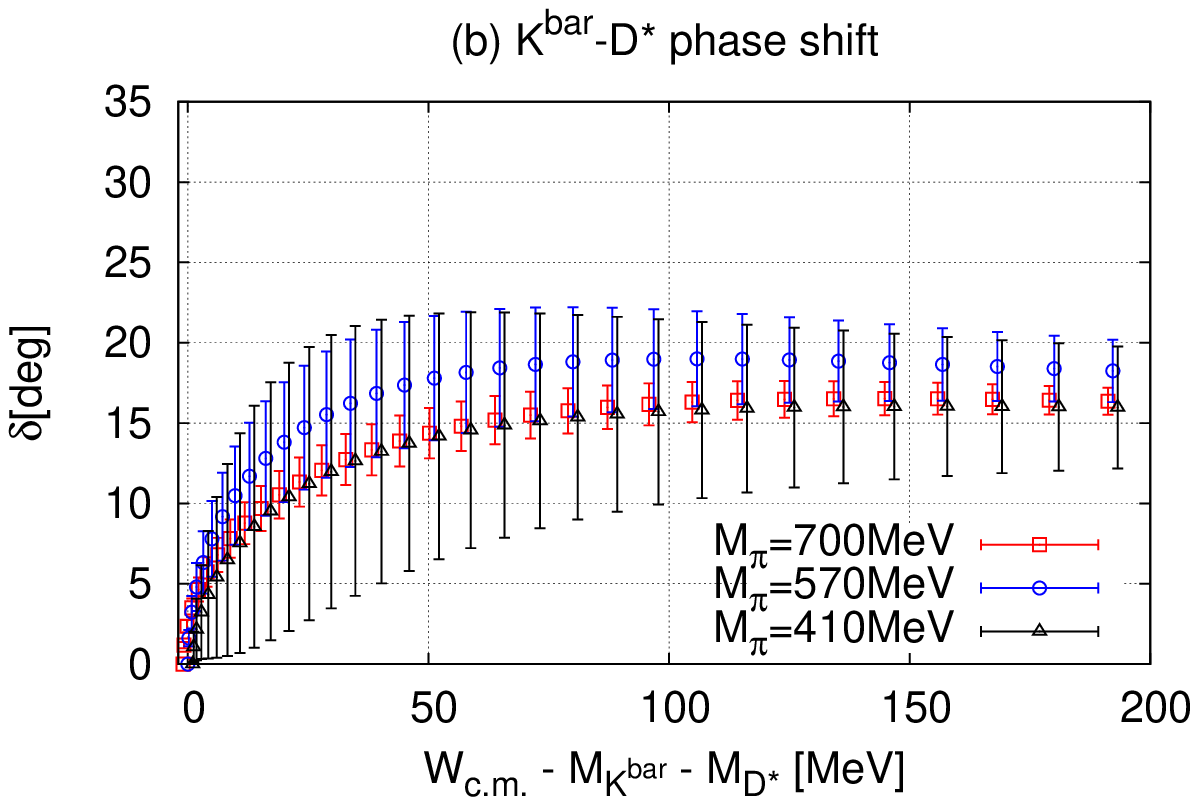}
\includegraphics[width=0.6\textwidth]{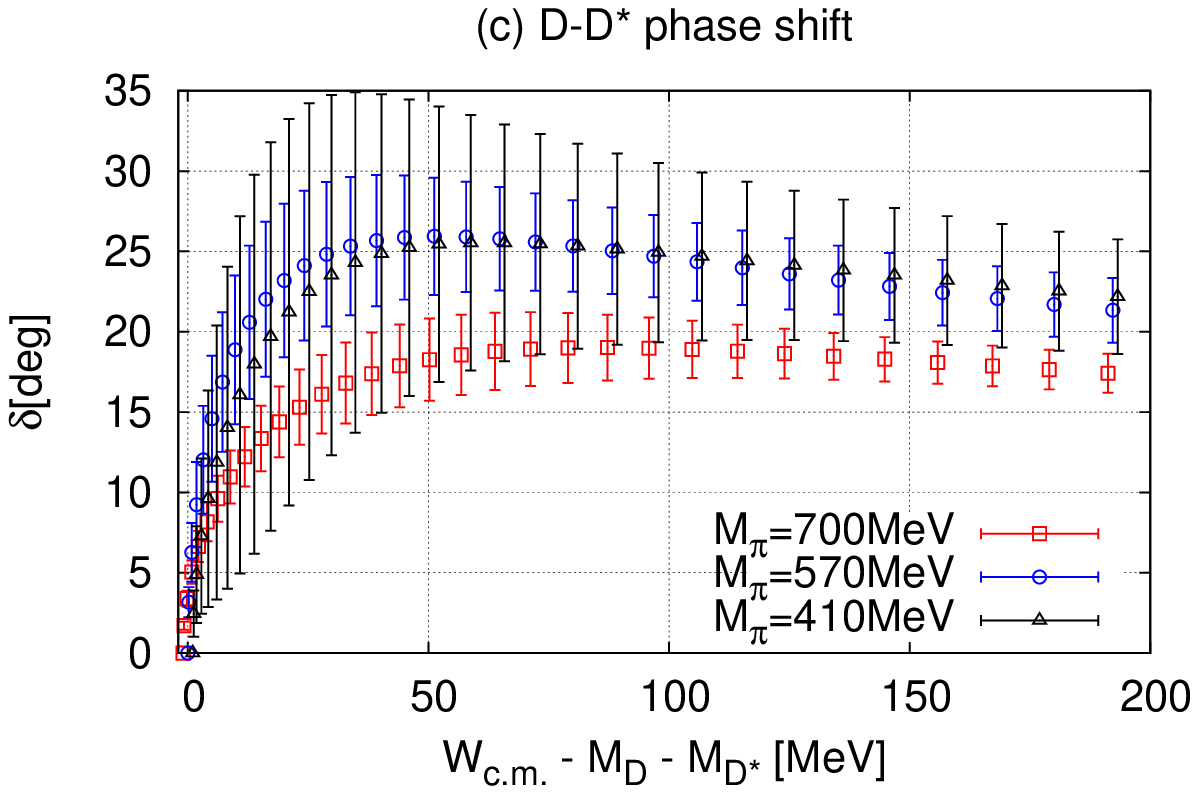}
\end{center}
\caption{
(color online)
S-wave $I=0$ scattering phase shifts in the (a) $\bar{K}$-$D$, (b) $\bar{K}$-$D^{*}$ and (c) $D$-$D^{*}$ channels.
Vertical error bars represent both statistical and systematic errors.
}
\label{fig3}
\end{figure}
\begin{figure}[htbp]
\begin{center}
\includegraphics[width=0.7\textwidth]{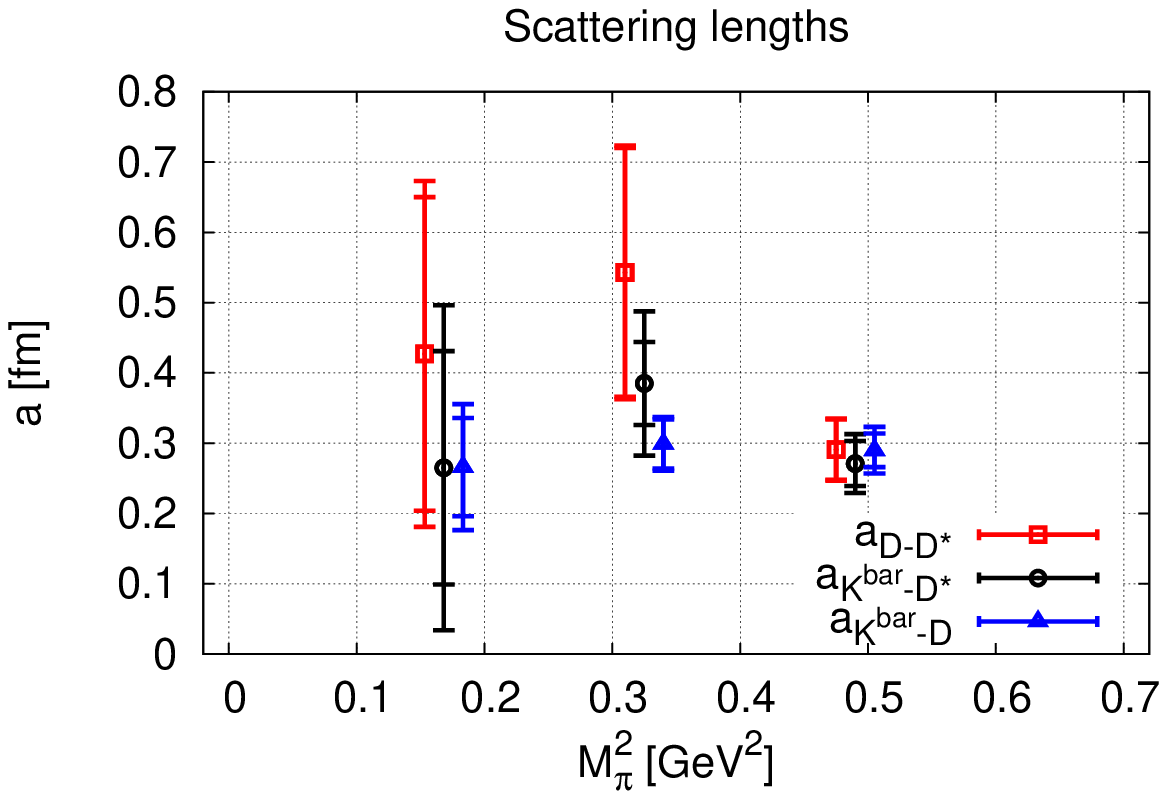}
\end{center}
\caption{
(color online)
Pion mass dependences of scattering lengths in the $I=0$ $\bar{K}$-$D$, $\bar{K}$-$D^{*}$ and $D$-$D^{*}$ channels.
Inner vertical error bars represent statistical errors.
Both statistical and systematic errors are included in total error bars.
}
\label{fig4}
\end{figure}

To investigate the possible existence of bound states or resonances in the $I=0$ channels,
we fit the potentials 
with analytic functions of $r$ and
solve the Schr\"{o}dinger equation with the fitted potentials at a given time-slice.
We employ a multi-range Gaussian form to  fit the potential, namely 
$g(r) \equiv \sum_{n=1}^{N_{\rm max}} V_n \cdot \exp (-\nu_n r^2)$ with 
$V_n$ and $\nu_n$ being fit parameters.
For all cases, we obtain good fits with  $\chi^2/{\rm d.o.f.} \simeq 0.6$ for $N_{\rm max}=4$.
Repeating this at different time-slices,
the mean values of scattering phase shifts are obtained from the weighted average over the time-slices $t/a=13$ through $18$.
Statistical errors for the scattering phase shifts are calculated by the jackknife method, 
and systematic errors are evaluated by the difference between the weighted average of the phase shifts over the time-slices $t/a=13-15$ and $t/a=16-18$.

Fig.~\ref{fig3} shows the resultant S-wave scattering phase shifts
as a function of the meson-meson center-of-mass energy 
in the $I=0$ $\bar{K}$-$D$ ($J^P=0^+$), $\bar{K}$-$D^{*}$ ($J^P=1^+$)
and $D$-$D^{*}$ ($J^P=1^+$) channels.
In Table~\ref{tab2} and Fig.~\ref{fig4}, we give the corresponding scattering lengths.
We do not find the negative-energy eigenvalues corresponding to the bound state solutions by solving the Schr\"{o}dinger equation
with the potentials shown in Fig.~\ref{fig2}.
Fig.~\ref{fig3} also indicates that there are no
bound states or resonances in this range of pion masses, $m_{\pi}=410 \sim 700$ MeV.
Although the potentials for  $D$-$D^{*}$ and $\bar{K}$-$D^{*}$ are not so much different,
as seen in the right panels of Fig.~\ref{fig2},
the scattering length in the $D$-$D^{*}$ channel is larger than that in the $\bar{K}$-$D^{*}$ channel.
This can be attributed to the smaller kinetic energy of $D$ 
in comparison to $\bar{K}$ due to a heavier charm quark.
A similar tendency has also been reported in studies of phenomenological models (see e.g. \cite{Yasui2009}).

Although we find  a good evidence of a sizable attraction in the $I=0$ channel at $m_{\pi}=410 \sim 700$ MeV, the existence of a bound or resonant $T_{cc} (J^P=1^+, I=0)$  at the physical point remains an open question\footnote{ If we take the same attractive
   potential as $D$-$D^{*}$ ($J^P=1^+,I=0$) at $m_{\pi}=410$MeV and calculate the  
    $B$-$B^{*}$ ($J^P=1^+,I=0$) channel with the physical masses of $B$ and $B^{*}$,
     we find a bound state with the binding energy 5.7(2.3)MeV.}.

\begin{table}[htbp]
   \centering
   \begin{tabular}{cccc}
      \hline
      \hline
$m_{\pi}$ (MeV) & 411(2) & 572(2) & 699(1)\\
      \hline 
$a_{\bar{K}D}$ (fm) & 0.266(70)(56) & 0.299(35)(15)  & 0.290(24)(23) \\
      \hline 
$a_{\bar{K}D^{*}}$ (fm)   & 0.265(166)(161) & 0.385(59)(84) & 0.271(32)(27) \\
      \hline 
$a_{DD^{*}}$ (fm)   & 0.427(223)(104) & 0.543(177)(35) & 0.291(43)(10) \\
      \hline
      \hline
   \end{tabular}
   \caption{ 
   Scattering lengths in  the $I=0$ channels for the
   $\bar{K}$-$D$, $\bar{K}$-$D^{*}$ and $D$-$D^{*}$ systems.
   The statistical and systematic errors are also shown.
   }
   \label{tab2}
\end{table}

\section{Summary}

 In order to clarify the possible existence of charmed tetraquark states ($T_{cc}$ and $T_{cs}$),  
 we have studied the S-wave meson-meson interactions
in several $I=0$ and $I=1$ channels ($D$-$D$, $\bar{K}$-$D$, $D$-$D^{*}$ and $\bar{K}$-$D^{*}$),
  using 
  (2+1)-flavor full QCD gauge configurations  generated at
 $m_{\pi} = 410 \sim 700$ MeV. For the charm quark, 
 we have employed the relativistic heavy-quark action  to take into account its dynamics on the lattice.

 S-wave meson-meson potentials are extracted from Nambu-Bethe-Salpeter wave functions
 using the HAL QCD method. Potentials are then used to calculate  scattering phase shifts and scattering lengths.
S-wave meson-meson interactions in the $I=1$ channels are found to be repulsive
and insensitive to the pion mass in the region we explored,
so that tetraquark  bound states are unlikely to be formed even at the physical pion mass.
On the other hand, the S-wave interactions in the $I=0$ channels show attractions 
in the $\bar{K}$-$D$, $\bar{K}$-$D^{*}$ and $D$-$D^{*}$ channels, which are 
qualitatively consistent with the phenomenological diquark picture.
S-wave scattering phase shifts in these attractive channels indicate, however,
that no bound states or resonances are formed at the pion masses used in this study, 
$m_{\pi} = 410 - 700$ MeV, though attractions become more prominent
as the pion mass decreases, particularly in the $I=0$ $D$-$D^{*}$ channel corresponding to
$T_{cc} (J^P=1^+, I=0)$. 

To make  a definite conclusion on the fate of $T_{cc}$ and $T_{cs}$ in the real world,
simulations near or at the physical point are necessary. 
We are planning to carry out such simulations with the PACS-CS (2+1)-flavor full QCD configurations with coupled-channel schemes~\cite{Aoki:2012bb,Sasaki2012}.

\section*{Acknowledgment}
The authors 
thank T. Hyodo, T. Matsuki, M. Oka, S. Takeuchi, M. Takizawa and S. Yasui
for fruitful discussions, and 
authors and maintainer of CPS++~\cite{CPS}, whose modified version is used in this Letter.
We also thank ILDG/JLDG~\cite{JLDG} for providing us with full QCD gauge configurations used in this study.
Numerical calculations were carried out on NEC-SX9 and SX8R at Osaka University and SR16000 at  YITP in Kyoto University.
This project is supported in part by by Grant-in-Aid
for Scientific Research on Innovative Areas(No.2004:20105001, 20105003) and
for Scientific Research ( Nos. 25800170, 25287046, 24740146,  24540273, 23540321) and SPIRE (Strategic Program for Innovative Research).
T.H. was partially supported by RIKEN iTHES project.




\begin{thebibliography}{99}
\bibitem{Jaffe1977}
  R.L.~Jaffe, Phys. Rev. Lett. {\bf 38} (1977) 195.


\bibitem{Zouzou:1986qh}
  S.~Zouzou, B.~Silvestre-Brac, C.~Gignoux and J.~M.~Richard,
  Z.\ Phys.\ C {\bf 30} (1986) 457.

\bibitem{Lipkin1986}
  H.J.~Lipkin, Phys. Lett. {\bf B 172} (1986) 242.

\bibitem{Carlson:1987hh}
  J.~Carlson, L.~Heller and J.~A.~Tjon,
  Phys.\ Rev.\ D {\bf 37} (1988) 744.

 \bibitem{Manohar1993}
  A.~V.~Manohar and M.~B.~Wise,
  Nucl.\ Phys.\ B {\bf 399} (1993) 17.



 \bibitem{DelFabbro:2004ta}
  A.~Del Fabbro, D.~Janc, M.~Rosina and D.~Treleani,
  Phys.\ Rev.\ D {\bf 71} (2005) 014008
  [hep-ph/0408258].
  
\bibitem{Cho:2010db}
  S.~Cho {\it et al.}  [ExHIC Collaboration],
  Phys.\ Rev.\ Lett.\  {\bf 106} (2011) 212001
  [arXiv:1011.0852 [nucl-th]].

   \bibitem{Hyodo2012}
  T.~Hyodo, Y.~-R.~Liu, M.~Oka, K.~Sudoh and S.~Yasui,
  Phys.\ Lett.\ B {\bf 721} (2013) 56.
\bibitem{Rashidin:2013voa}
  R.~Rashidin,
  arXiv:1305.5711 [hep-ph].
  
\bibitem{Selem:2006nd}
  A.~Selem and F.~Wilczek,
  hep-ph/0602128.

  
  \bibitem{Yasui2009}
  S.~H.~Lee and S.~Yasui,
  Eur.\ Phys.\ J.\ C {\bf 64} (2009) 283.

\bibitem{Carames:2011zz}
  T.~F.~Carames, A.~Valcarce and J.~Vijande,
  Phys.\ Lett.\ B {\bf 699} (2011) 291.

    \bibitem{Ohkoda2012}
  S.~Ohkoda, Y.~Yamaguchi, S.~Yasui, K.~Sudoh and A.~Hosaka,
  Phys.\ Rev.\ D {\bf 86} (2012) 034019, 
 and references therein.
  
\bibitem{Vijande:2013qr}
  J.~Vijande, A.~Valcarce and J.~-M.~Richard,
  Phys.\ Rev.\ D {\bf 87} (2013) 034040
  [arXiv:1301.6212 [hep-ph]]. 
  

  
\bibitem{Brown:2012tm}
  Z.~S.~Brown and K.~Orginos,
  Phys.\ Rev.\ D {\bf 86} (2012) 114506
  [arXiv:1210.1953 [hep-lat]].
  
\bibitem{Bicudo:2012qt}
  P.~Bicudo and M.~Wagner,
  arXiv:1209.6274 [hep-ph].
    

\bibitem{Ikeda:2011bs}
  Y.~Ikeda and H.~Iida,
  Prog.\ Theor.\ Phys.\  {\bf 128} (2012) 941
  [arXiv:1102.2097 [hep-lat]].

  \bibitem{Kawanai:2011xb}
  T.~Kawanai and S.~Sasaki,
  Phys.\ Rev.\ Lett.\  {\bf 107} (2011) 091601
  [arXiv:1102.3246 [hep-lat]].

  
\bibitem{RHQ_Aoki}
  S.~Aoki, Y.~Kuramashi and S.~-i.~Tominaga,
  Prog.\ Theor.\ Phys.\  {\bf 109} (2003) 383.



\bibitem{Ishii2007a}
N.~Ishii, S.~Aoki and T.~Hatsuda, 
Phys. Rev. Lett. {\bf 99} (2007) 022001.


\bibitem{Aoki2010}
S.~Aoki, T.~Hatsuda and N.~Ishii, 
 Prog. Theor. Phys. {\bf 123} (2010) 89.



\bibitem{Ishii2012}
  N.~Ishii {\it et al.}  [HAL QCD Collaboration],
  Phys.\ Lett.\ B {\bf 712} (2012) 437.

\bibitem{HAL2012}
  S.~Aoki {\it et al.}  [HAL QCD Collaboration],
  PTEP {\bf 2012} (2012) 01A105.


\bibitem{Kurth:2013tua}
  T.~Kurth, N.~Ishii, T.~Doi, S.~Aoki and T.~Hatsuda,
  arXiv:1305.4462 [hep-lat].

\bibitem{Miransky}
  V.~A.~Miransky, {\it DYNAMICAL SYMMETRY BREAKING IN QUANTUM FIELD THEORY} 
  (World Scientific Publishing, 1993), Chap 8.


\bibitem{Aoki:2012bb}
  S.~Aoki, B.~Charron, T.~Doi, T.~Hatsuda, T.~Inoue and N.~Ishii,
  Phys.\ Rev.\ D {\bf 87} (2013) 3,  034512
  [arXiv:1212.4896 [hep-lat]].
  


\bibitem{PACS-CS2009}
  S.~Aoki {\it et al.}  [PACS-CS Collaboration],
  Phys.\ Rev.\ D {\bf 79} (2009) 034503.

\bibitem{PACS-CS2010}
  S.~Aoki {\it et al.}  [PACS-CS Collaboration],
  Phys.\ Rev.\ D {\bf 81} (2010) 074503.

\bibitem{Namekawa2011}
  Y.~Namekawa {\it et al.}  [PACS-CS Collaboration],
  Phys.\ Rev.\ D {\bf 84} (2011) 074505.


\bibitem{Sasaki2012}
  K.~Sasaki [HAL QCD Collaboration],
  PoS LATTICE {\bf 2012} (2012) 157.


\bibitem{CPS}
Columbia Physics System (CPS), http://qcdoc.phys.columbia.edu/cps.html

\bibitem{JLDG}
International Lattice Data Grid,
http://www.lqcd.org/ildg;
Japan Lattice Data Grid,
http://www.jldg.org

\end{thebibliography}
\end{document}